\documentclass{aa501}
\usepackage{graphicx}

\begin{document}

\newcommand{\inthms}[3]{$#1^{\rm h}#2^{\rm m}#3^{\rm s}$}
\newcommand{\dechms}[4]{$#1^{\rm h}#2^{\rm m}#3\mbox{$^{\rms}\mskip-7.6mu.\,$}#4$}
\newcommand{\intdms}[3]{$#1^{\circ}#2'#3''$}
\newcommand{\decdms}[4]{$#1^{\circ}#2'#3\mbox{$''\mskip-7.6mu.\,$}#4$}
\newcommand{\tmb}{\mbox{T$_{\rm mb}$}}
\newcommand{\OI}{\mbox{O\,{\sc i}}}
\newcommand{\dtco}{D$_{2}$CO}
\newcommand{\hdco}{HDCO}
\newcommand{\htco}{H$_{2}$CO}
\newcommand{\httco}{H$_{2}^{13}$CO}
\newcommand{\kmps}{km s$^{-1}$}
\newcommand{\eg}{\mbox{\hbox{\it e.g.,}}}

\hyphenation{Fe-bru-ary Gra-na-da mo-le-cu-le mo-le-cu-les}


\title{Discovery of three nearby L dwarfs in the Southern Sky 
\thanks{Based on observations collected with the ESO\,3.6m/EFOSC2 and the NTT/SOFI
at the European Southern Observatory, La Silla, Chile
(ESO programme 68.C-0664).}}

\author{N. Lodieu \inst{} \and
       R.-D. Scholz \inst{} \and
       M. J. McCaughrean \inst{}}
\institute{Astrophysikalisches Institut Potsdam, An der Sternwarte 16, 14482 Potsdam, Germany}

\offprints{Nicolas Lodieu, nlodieu@aip.de}

\date{Received 19 April 2002 / Accepted }

\titlerunning{Discovery of three L dwarfs in the Southern Sky}
\authorrunning{Lodieu et al.}
\abstract{ We report the discovery of three L dwarfs in the solar vicinity within
30 parsecs.
These objects were originally found as proper motion objects from a combination of $R$
and $I$ photographic plates measured as part of the SuperCOSMOS Sky Surveys.
We subsequently identified these objects as bona fide
brown dwarf candidates on the basis of their $R$-$I$ colour, as first criterion,
and subsequently their $J$-$K$ colours when the infrared data were 
available from the 2MASS database. 
Spectroscopic observations in the optical with the 
ESO\,3.6m/EFOSC2 and in the near-infrared with the NTT/SOFI led to the
classification of their spectral types as early L dwarfs.}

\maketitle

\keywords{surveys -- stars: distances -- stars: kinematics -- stars: late-type --
stars: low-mass, brown dwarfs -- solar neighbourhood
}

%
%

\section{Introduction}
The first L dwarf, Kelu-1, was discovered in the solar neighbourhood in 1997
(Ruiz, Leggett \& Allard ~\cite{ruiz97}). Up to now, more
than 100 L dwarfs have been discovered mostly by DENIS, 2MASS, and SDSS all-sky 
surveys via optical spectroscopy based on their optical/infrared colours
(Delfosse et al.~\cite{delfosse97}; Delfosse et al.~\cite{delfosse99}; 
Kirkpatrick et al.~\cite{kirkpatrick99}; Reid et al.~\cite{reid00};
Fan et al.~\cite{fan00}; Schneider et al.~\cite{schneider02} and references therein).
According to theoretical models (Burrows et al.~\cite{burrows97}, 
Baraffe et al.~\cite{baraffe98}), most of them are expected to be brown dwarfs 
with effective temperature below 2000\,K and ages of few Gyr. The substellar status 
of some late M and L dwarfs has been confirmed with subsequent lithium detections
(Tinney, Delfosse, \& Forveille~\cite{tinney97};
Delfosse et al.~\cite{delfosse97}; Tinney et al.~\cite{tinney98};
Delfosse et al.~\cite{delfosse99}; Mart\'{\i}n et al.~\cite{martin99};
Fan et al.~\cite{fan00}).
All these objects are nearby, within 50 pc as shown by parallax measurements
and, therefore, exhibit significant proper motions. Three years ago,
Scholz et al.~(\cite{scholz00}) started a new high proper motion survey in the
Southern Sky using UK Schmidt plates spanning roughly 10--20 years. The
APM (Automatic Plate Measuring) photographic colours ($B_{J}$ and $R$)
have proven a useful discriminant of nearby low-mass objects straddling
the stellar/substellar boundary (Kirkpatrick et al.~\cite{kirkpatrick97}).
Our selection criterion for very red and high proper motion objects provides a powerful 
tool for detecting low-luminosity objects, and, in this letter, we report
the discovery of three early L dwarfs found via low-dispersion optical 
and near-infrared spectroscopy using ESO\,3.6m/EFOSC2 and NTT/SOFI.

\section{Sample selection}
Brown dwarf candidates have been selected among extremely red objects
discovered in different new high proper motion surveys in the southern
sky. A detailed description of the different search methods and samples will
be given in a forthcoming paper, but all are based on Schmidt
plates scanned with various 
measuring machines. 
The three objects mentioned in this
study were discovered from SuperCOSMOS Sky Surveys(SSS) data
(Hambly et al.~\cite{hambly01a}; Hambly, Irwin \&
MacGillivray~\cite{hambly01b}; Hambly et al.~\cite{hambly01c}).
$B_{J}$, $R$ and 
$I$ band measurements of UKST plates with additional $R$ band
data from ESO Schmidt plates have been combined to search for proper motion
objects.

The basic search strategy consisted of looking for objects
on a given plate which were not matched with a corresponding
object in a different passband to within a nominal search
radius of 3 arcsec. This process was repeated for each available
plate at a given location on the sky. These reduced catalogues 
of unmatched objects were then compared, looking for possible
counterparts out to a search radius of 1 arcmin. Objects were
identified as proper motion candidates if they were picked up
at least three times along a straight line, to within errors
on the implied proper motion of 30 mas/year. Later, further
positional information from (\eg{}) the 2MASS survey was used
to refine these detections. 
The epoch differences between the plates
vary from field to field, with typically a maximum of
15--20 years between the UKST $B_J$ and $R$ plates.
The search was therefore sensitive to proper motions larger than 0.15--0.20
arcsec/yr.

As all three objects were discovered in SSS data, they have correspondingly
been named SSSPM\,J0219$-$1939, SSSPM\,J2310$-$1759, and SSSPM\,J0109$-$5101.
The first two objects were also identified on the POSS-I plates also measured
as part of the SSS (although not on the UKST $B_J$ plates), and in the 2MASS
2nd incremental release public data base. No 2MASS measurements are presently
available for SSSPM\,J0109$-$5101, and thus its $K_s$ magnitude was derived
from the SOFI acquisition image.
Table~\ref{sss2mass} lists the data obtained from SSS and 2MASS\@.
The proper motions of the first two objects given in Table~\ref{sss2mass}
are based on the positions on two $R$ and one $I$ band plates, as well as on
the additional SSS measurements of the POSS-I plates and on the 2MASS positions.
The proper motion of SSSPM~J0219$-$1939 was obtained from $B_J$, $R$, and
$I$ band measurements of UKST plates. Positions are given at the latest
available epoch. 

\begin{table*}
 \caption[]{Astrometry and photometry from SSS and 2MASS\@. 2MASS infrared magnitudes 
are not available for SSSPM~J0109$-$5101. 
$^{a}$ K$_{s}$ photometry was computed from the SOFI acquisition images.}
 \label{sss2mass}
 \begin{tabular}{cccrcccccc}
 \hline
Name     & $\alpha, \delta$ & Epoch & $\mu_{\alpha}\cos{\delta}$ & $\mu_{\delta}$ & $R$ & $I$ & $J$ & $H$ & $K_s$ \\
SSSPM~J..& (J2000) & & \multispan{2}{\hfil mas/yr \hfil} & \multispan{2}{\hfil (SSS) \hfil} & \multispan{3}{\hfil (2MASS) \hfil} \\
 \hline
0219$-$1939& 02 19 28.03 $-$19 38 41.0 &1999.87& $+194 \pm 04$ & $-173 \pm 06$ & 20.13 &     17.46 &14.09 &13.30 &12.83  \\
2310$-$1759& 23 10 18.53 $-$17 59 09.4 &1998.10& $ +24 \pm 17$ & $-246 \pm 13$ & 20.52 &     17.67 &14.40 &13.58 &13.01  \\
0109$-$5101& 01 09 01.29 $-$51 00 51.1 &1990.80& $+207 \pm 04$ & $ +94 \pm 11$ & 18.21 &     14.80 &      &      &10.96 $^{a}$ \\
 \hline
 \end{tabular}
\end{table*}

\section{Observations and data reduction}
\subsection{ESO3.6m/EFOSC2}
\begin{figure}[htb]
\begin{center}
\includegraphics[width=\columnwidth, angle=0]{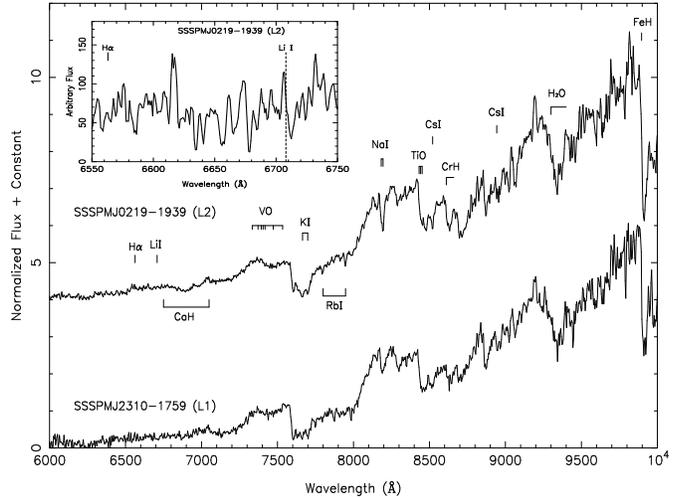}
\caption{Low-resolution (R $\sim$ 600) optical spectra of SSSPM~J0219$-$1939 (top) 
and SSSPM~J2310$-$1759 (bottom) using the Grism 12 on the 3.6m/EFOSC2.
An arbitrary constant has been added to separate the spectra.
Their spectral types are estimated  to be L2 and L1, respectively (see text). 
Labelled are the TiO and VO 
absorption bands, alkali elements and hydride bands, as well as the locations of
lithium at 6708 \AA{} and H$\alpha$ at 6563 \AA{}.}
\label{optsp}
\end{center}
\end{figure}

Spectroscopy was first obtained in the optical for 
SSSPM~J0219$-$1939 and SSSPM~J2310$-$1759 with EFOSC2 mounted on the ESO\,3.6m 
telescope
at La Silla on 22 November 2001. The night was photometric 
and the seeing around
0.6--0.8 arcsec. The camera uses a 2048 $\times$ 2048 pixels 
Loral/Lesser CCD. 
The pixel size is 0.157 arcsec and
the useful field of view is 5.2 $\times$ 5.2 arcmin.
A 1 arcsec slit has been used for spectroscopy with Grism 12
covering 6000--10300 \AA{}, yielding a resolution of $\sim$ 600.
Three exposures of 1200 seconds each, shifted
along the slit by 30 arcsec have been obtained for both candidates. 
The data reduction consisted in subtracting an averaged dark frame 
and dividing by an internal quartz flat field taken just after the first exposure 
of the target in order to remove efficiently fringing above 8000 \AA{}.
Wavelength calibration 
was made using He and Ar lines throughout the whole wavelength range. Flux calibration 
was achieved using an averaged sensitivity function determined with several exposures of
the spectrophotometric standard stars 
EG21 and LTT1788 during the night.
The co-added optical spectra of SSSPM~J0219$-$1939 and SSSPM~J2310$-$1759 
are shown in Fig.~\ref{optsp}. The TiO and VO absorption
bands and ground-state transitions of the alkali elements including 
Na\,{\small{I}} and K\,{\small{I}} doublets, 
are labelled on each spectrum, along with the locations
of the lithium and H$\alpha$ lines at 6708 \AA{} and 6563 \AA{}, respectively. 
Furthermore, a higher resolution spectrum of the lithium and H$\alpha$ region of 
SSSPM~J0219$-$1939 has been obtained with the Grism 10 (6280--8200 \AA{})
on ESO\,3.6m/EFOSC2 in January 2002 yielding a dispersion of 1 \AA{}.
Data reduction was standard and similar to the one applied for Grism 12
described previously. A zoom of the lithium and H$\alpha$ region
of SSSPM~J0219$-$1939 is plotted in the left upper corner of Fig.~\ref{optsp}.

\subsection{NTT/SOFI}
\begin{figure}[htb]
\begin{center}
\includegraphics[width=0.7\columnwidth, angle=270]{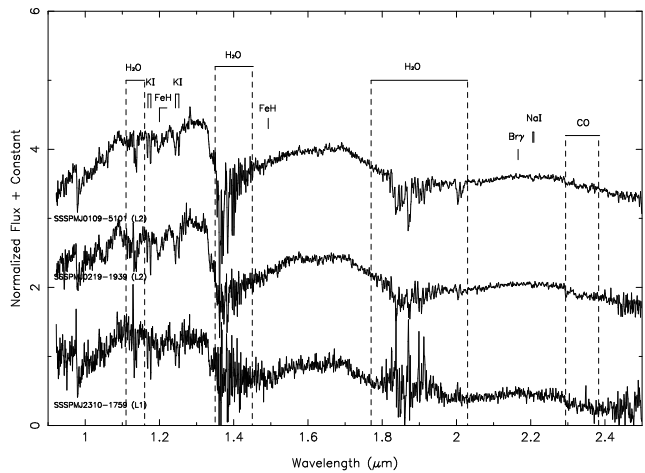}
\caption{Near-infrared spectra of SSSPM~J2310$-$1759, SSSPM~J0219$-$1939 and
SSSPM~J0109$-$5101 from bottom to top. An arbitrary constant
has been added to separate the spectra. The locations of prominent bands of H$_{2}$O, 
FeH and CO are labelled, as well as atomic lines including K\,{\small{I}} 
and Na\,{\small{I}}.} 
\label{irsp}
\end{center}
\end{figure}

Initial assessment of our optical spectra immediately identified these two sources
as early L dwarfs. SSSPM~J0219$-$1939 and SSSPM~J2310$-$1759 have been
followed up at longer wavelengths (0.95--2.5 $\mu$m) with the 
near-infrared camera/spectrograph SOFI mounted on
the New Technology Telescope at La Silla on 24 November 2001. 
SOFI is equipped with a 1024 $\times$ 1024 pixels HgCdTe (HAWAII) array
(Moorwood \& Spyromilio~\cite{moorwood}) with a pixel size of 0.294 arcsec for 
the Large Field Objective used for spectroscopy. The third object in our present
sample, SSSPM~J0109$-$5101, was selected on the basis of its $R-I$ colour alone
and observed on 25 November 2001 with the same instrument configuration.
Both NTT nights were photometric and the seeing variable from 0.8--1.0 arcsec.
A 1 arcsec slit was used yielding R $\sim$ 600 for both blue (0.95-1.64 $\mu$m)
and red (1.53-2.52 $\mu$m) gratings. Featureless spectrophotometric standards
(typically F5--F7) were measured within 1 degree on the sky to remove telluric 
absorption and flux calibrate. Three positions along the slit have been made
with 250 seconds each for SSSPM~J0219$-$1939 and SSSPM~J2310$-$1759 and 60 seconds
each for SSSPM~J0109$-$5101 to subtract the sky. Each individual frame has been
flat-fielded, sky-subtracted and a one-dimensional spectrum extracted. This
spectrum has then been divided by the spectrophotometric standard and multiplied
by a template smoothed to our resolution, whose spectral type is identical
to the observed standard. Infrared spectra for the three L dwarfs are shown in 
Fig.~\ref{irsp}, 
along with the locations of H$_{2}$O and CO bands and atomic lines. 

\section{Results and discussion}
\subsection{Spectral type classification}
The optical spectra of SSSPM~J0219$-$1939 and SSSPM~J2310$-$1759 reveal
features typical of early L dwarfs. The TiO and VO 
absorption bands are weaker than those found in M dwarfs. K\,{\small{I}} 
and Na\,{\small{I}} doublets 
are also clearly seen although the latter is barely resolved at this low
resolution. Hydride bands including CrH, FeH and CaH as well as other alkali 
elements including Cs\,{\small{I}} and Rb\,{\small{I}} clearly emerge 
from the optical spectra shown in Fig.~\ref{optsp}. 
Furthermore, H$\alpha$ is weakly seen in both objects which probably means that
our targets were observed in a quiet state since a H$\alpha$ flare
has been recently reported in a L3 dwarf (Hall\ ~\cite{hall02}).
Moreover, a higher resolution spectrum of SSSPM~J0219$-$1939 has been 
obtained in January 2002 using the Grism 10 (6280--8200 \AA{}) on
ESO\,3.6m/EFOSC2 to search for lithium at 6708 \AA{} (see Fig.~\ref{optsp}).
However, no significant absorption line is detected at this location
which set the lower mass limit of this object at 0.065 M$_{\odot}$.
After the discovery of several objects cooler than M dwarfs, two spectral 
type schemes have been defined by 
Kirkpatrick et al.~(\cite{kirkpatrick99}) and  Mart\'{\i}n et al.~(\cite{martin99})
to classify late M and L dwarfs. On the one hand, 
Kirkpatrick et al.~(\cite{kirkpatrick99})
considers the median spectral type value computed for
several ratios of alkali elements and oxide absorption bands (see 
Table~\ref{spind}).
On the other hand, the PC3 index is found to be the most reliable index 
among those defined by Mart\'{\i}n et al.~(\cite{martin99})
to classify M and L dwarfs. Computations of the different ratios
mentioned above led to consistent results for SSSPM~J0219$-$1939
and SSSPM~J2310$-$1759 whose spectral types have been estimated to be
L2 and L1, respectively.

Infrared spectra of the three objects quoted in this study denote strong
H$_{2}$O absorption bands, the CO break around 2.3$\mu$m, and strong atomic 
lines including K\,{\small{I}}, features characteristic of L dwarfs 
(see Fig.~\ref{irsp}).
Recently, several infrared indices have been defined mostly based on the strength
of water absorption bands (Mart\'{\i}n et al.~\cite{martin99},
Tokunaga \& Kobayashi~\cite{tokunaga99}, Reid et al.~\cite{reid01},
Testi et al.~\cite{testi01}, Geballe et al.~\cite{geballe02},
Mart\'{\i}n~\cite{martin00}). The name of each index is given in 
Table~\ref{spind} 
and the definition refers to the papers quoted above. 
Most of the spectral type computations are internally consistent yielding L1
except the one defined by Tokunaga \& Kobayashi~\cite{tokunaga99} and
Reid et al.~\cite{reid01} where discrepancies are noticed. As
the optical classifications are, up to now, more accurate than
the infrared schemes, spectral types of L1 and L2 are kept for
SSSPM~J2310$-$1759 and SSSPM~J0219$-$1939, respectively. In the case of
SSSPM~J0109$-$5101, comparison of the infrared spectra plotted in 
Fig.~\ref{irsp} shows that 
this object is later than SSSPM~J2310$-$1759 and similar to SSSPM~J0219$-$1939.
A spectral type of L2 is therefore adopted with an uncertainty of a
subclass.
Meanwhile, the analysis of the $K$ band spectra via two indices defined by
Tokunaga \& Kobayashi~(\cite{tokunaga99}) provide a rough estimate of the 
effective temperature spanning 1800--2000\,K for our objects. 
As the authors claim that brown dwarfs are likely to have
K1 $\geq$ 0.250, the computed values for our objects are clearly below
and therefore they are probably not brown dwarfs although SSSPM~J0109$-$5101 has the 
highest probability to lie in the substellar regime. A high-resolution 
optical spectrum of this object is required to check this assessment.

\begin{table}[htb]
\begin{center}
 \caption[]{Optical and near-infrared indices and
corresponding spectral type for the three candidates
(ST). Letters in brackets refer to the following
references. Mart\'{\i}n et al.~1999 [a],
Kirkpatrick et al.~1999 [b],
Tokunaga \& Kobayashi 1999 [c], Reid et al.~2001 [d], Testi et al.~2001 [e],
Geballe et al.~2001 [f] and Mart\'{\i}n 2000 [g]}
 \label{spind}
 \begin{tabular}{lccc}
 \hline
Ref \& Feature & Value (ST)  & Value (ST) & Value (ST) \\
SSSPM\,J.. &  0219$-$1939 & 2310$-$1759 & 0109$-$5101 \\
 \hline
[a] PC3                  & 3.05 (L2)   & 2.30 (L1) &     \cr
[b] CrH-a                & 1.36 (L1)  & 1.17 (L0) &      \cr
[b] Rb-b/TiO-b           & 1.10 (L2--L3)  & 0.47 (?)  &   \cr
[b] Cs-a/VO-b            & 1.13 (L3) & 0.99 (L2) &       \cr
\hline
[c] K1                   & 0.158 (L1) & 0.128 (L0)  & 0.223 (L2) \cr
[c] K2                   & $-$0.01    & $-$0.060    & $-$0.010    \cr
[d] H$_{2}$O$^{\rm A}$   & 0.60 (L4) & 0.72 (L0) & 0.66 (L2) \cr
[d] H$_{2}$O$^{\rm B}$   & 0.69 (L3) & 0.78 (L1) & 0.77 (L1.5) \cr
[e] sHJ                  & 0.45 (L1) & 0.35 (L1) & 0.33 (L1) \cr
[e] sKJ                  & 0.91 (L1) & 0.94 (L1) & 0.79 (L1) \cr
[e] sH$_{2}$O$^{\rm J}$  & 0.18 (L1) & $-$0.07 (L1) & 0.21 (L1) \cr
[e] sH$_{2}$O$^{\rm H1}$ & 0.40 (L1) & 0.22 (L1) & 0.29 (L1) \cr
[e] sH$_{2}$O$^{\rm H2}$ & 0.16 (L1) & 0.39 (L1) & 0.36 (L1) \cr
[e] sH$_{2}$O$^{\rm K}$  & 0.16 (L1) & 0.14 (L1) & 0.18 (L1) \cr
[f] H$_{2}$O 1.5$\mu$m   & 1.49 (L1) & 1.25 (L1) & 01.30 (L1) \cr
[g] Q$_{\rm H-band}$     & 0.41 (L1) & 0.37 (L2) & 0.42 (L1) \cr
\noalign{\smallskip}
 \hline
 \end{tabular}
\end{center}
\end{table}

\subsection{Distance estimates}
%
%
Assuming the spectral type determined above, distances can be computed using fiducial 
L dwarfs with known absolute 
magnitudes and trigonometric parallaxes (Kirkpatrick et al.~\cite{kirkpatrick00}).
Distances of 26.7 pc and 30.8 pc have been computed for 
SSSPM~J2310$-$1759 and SSSPM~J0219$-$1939, respectively. 
Photometry of SSSPM~J0109$-$5101 from the acquisition images yields K$_{s}$=10.96, 
roughly one magnitude brighter than Kelu-1, provides a distance of 13 pc assuming a distance
of 19.2 pc for Kelu-1 (Kirkpatrick et al.~\cite{kirkpatrick00}).
Then, the tangential velocities of these objects have been computed 
using classical formulae and velocities of 14.0 km/s$^{-1}$, 32.6 km/s$^{-1}$ and 
40.3 km/s$^{-1}$ derived for SSSPM~J0109$-$5101, SSSPM~J2310$-$1759, and SSSPM~J0219$-$1939,
respectively. These velocities are consistent with the kinematics 
of disk stars.


\section{Conclusions}

Three nearby L dwarfs have been discovered in a new high proper motion survey 
of the Southern Sky.
These objects are members of a larger sample of red and nearby proper motion candidates
which will be described in a forthcoming paper. Spectral types have been estimated
to be L1, L2 and L2 for SSSPM~J2310$-$1759, SSSPM~J0219$-$1939, and SSSPM~J0109$-$5101,
respectively. Optical and infrared classification schemes led to consistent results
within a subclass although some discrepancies are noticed depending on the
index used.
These three L dwarfs lie within 30 pc and have velocities consistent with disk
components. Further observations are required, first,  to improve distance estimates 
and, second, to assess their masses and thus potential brown dwarf status via
high-resolution spectroscopy enabling lithium detection.

\begin{acknowledgements}
The authors acknowledge enriching discussion with the support teams during
the different runs. We also thank Martin K\"urster, George Hau and
Gaspare Lo-Curto for taking the higher resolution spectra of SSSPM~J0219$-$1939.
The discovery of the faint high proper motion objects is based on
the SuperCOSMOS Sky Surveys, i.e. digitized
data obtained from scans of UKST and ESO
Schmidt plates. We would like to thank the SuperCOSMOS team for producing
such excellent data.
This research has also made use of data products
from the Two Micron All Sky Survey, which
is a joint project of the University of Massachusetts and the Infrared
Processing and Analysis Center, funded by the National Aeronautics and
Space Administration and the National Science Foundation.
NL thanks the European Research Training Network on ``The Formation and Evolution of Young 
Stellar Clusters'' (HPRN-CT-2000-00155) for financial support.
\end{acknowledgements}

\end{document}